\documentclass[thmsa,12pt,onecolumn]{article}
\usepackage{amsfonts}

\usepackage{sw20lart}

\input{tcilatex}
\usepackage{graphicx}

\begin{document}

\title{Glass transitions of freely suspended \\
polymer films}
\author{P.\ G.\ de Gennes \\
Coll\`{e}ge de France, 11 place M.\ Berthelot,\\
75231 Paris Cedex 05, France}
\maketitle

\begin{abstract}
We present a modified version of our ''sliding model'', where chain arcs,
between two contacts at the surface, may move if all the barriers along the
arc are weaker than a certain threshold. An important advance of the revised
model is that the high limiting chain lengths $N^{\ast }$ observed in the
experiments ($N^{\ast }\sim 10^{4}$) are naturally accounted for. 

In this model, a film of thickness $h<2R_{0}$ (where $R_{0}$ is the coil
size) can show either a ''sandwich'' structure with two mobile sublayers (at
low temperatures $T<T(h)$), or a single mobile layer at $T>T(h)$. But the
accident occurring at $T=T(h)$, does not necessarily coincide with the 
\textit{apparent }glass transition $T_{a}(h),$ determined by the
intersection of two tangents in a plot of thickness versus temperature.
\end{abstract}

\section{General aims}

Thin films of atactic polystyrene (of thickness 10-100 nanometers), freely
suspended in air, show an extraordinary depletion of their glass transition
point \cite{forrest 96}\cite{forrest 97}\cite{forrest 00}. Here, $T_{g}$ is
derived mainly from ellipsometric measurements of the thickness $h(T)$: the
low temperature expansion coefficient is smaller than the high temperature
coefficient, and they define on the plot $h(T)$ two tangents which intersect
at one temperature $T_{a}(h).$ It is assumed that the glass transition point
of the film coincides with $T_{a}.$

The depletion of $T_{g}$ occurs mainly for high molecular weights (above 500
K) but saturates at extremely high weights (above 2 000 K).\ The thicknesses
involved are comparable to the coil size: this induced us to propose that
''melting'' can occur via two competing scenarios: the first scenario
corresponds to the standard bulk transition, related to the
freezing-unfreezing of certain local degrees of freedom. The second scenario
involves sliding motions of each polymer chain along its own path \cite{de
gennes}.\ In the bulk, this sliding is blocked at the end points.\ But, near
a free surface, we expect that a thin ''skin'' (of thickness $\stackrel{\sim 
}{<}$ 1 nanometer) is fluid, and that the chain arcs which touch this skin
at two points A, B, may be fluidized in between.

The fluidity which is described here is \textit{not complete}: each chain is
still trapped between its two ends. This agrees with the observation that
films below the bulk glass transition point $T_{gb}$, but above the measured
''transition point'' $T_{a}(h)$, can be kept permanently while films at $%
T>T_{gb}$ generate holes which grow indefinitely (to decrease the surface
energy).

The sliding model leads naturally to some of the observed features \cite{de
gennes}.\ But it is not complete \ \ a) the nature of the sliding motion is
unclear: in its simplest version, it would be a simultaneous cooperative
motion of all monomers along a long arc AB, and this is unrealistic \ \ b)
the drops in $T_{g}$ were expected to be very large also for relatively
short chains (M$<500K$) while, in fact, they are found to be weak in this
limit.

In the present note, we return to two questions: \ \ a) in section 2, we
present a modified set of equations describing the onset of sliding which
does not require a cooperative motion of all the arc \ \ b) we argue that
the characteristic temperature $T_{a}(h)$, derived from the experiments by
the construction of fig. 1, need not be the actual transition point $T(h)$
(fig. 2).

\FRAME{fhFU}{2.1465in}{1.3439in}{0pt}{\Qcb{{\protect\small The standard
construction defining an apparent glass temperature T}$_{a}${\protect\small %
(h) from the plot h(T) of film thickness versus temperature.}}}{\Qlb{1}}{%
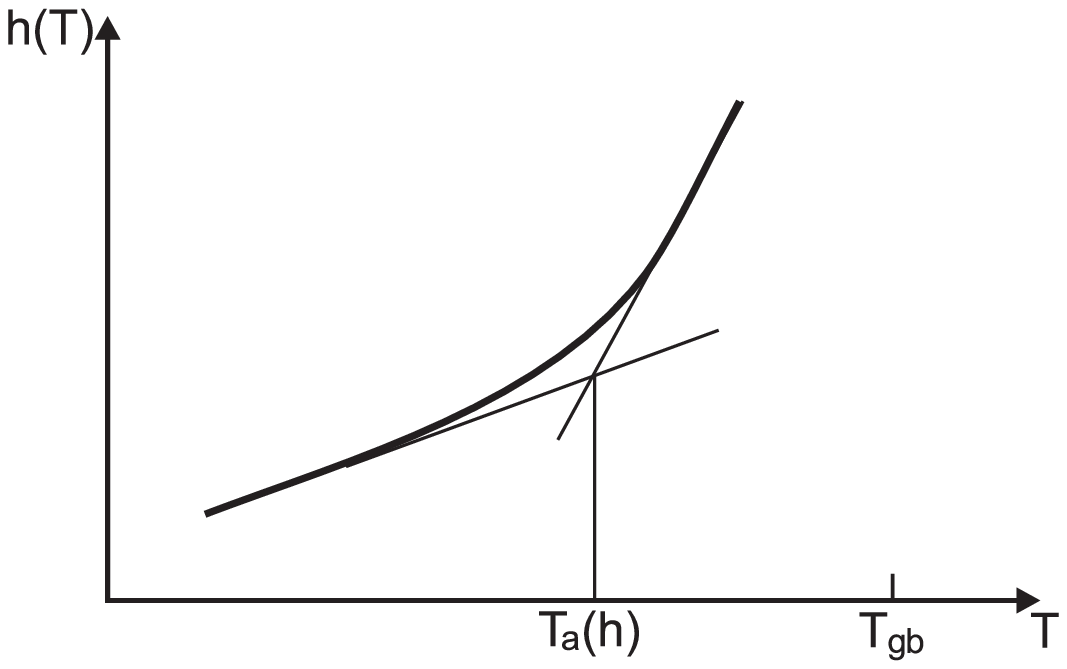}{\raisebox{-1.3439in}{\includegraphics[height=1.3439in]{fig11.eps}}}

\FRAME{fhFU}{2.2502in}{3.0536in}{0pt}{\Qcb{{}{\protect\small A tentative
plot of thickness versus temperature for the sliding model, in the regime
where h\TEXTsymbol{<}2R}$_{0}${\protect\small \ (R}$_{0}=${\protect\small \
coil size). a) at high temperature (T\TEXTsymbol{>}T}$_{gb}${\protect\small %
) the expansion coefficient (}$\protect\alpha _{L}${\protect\small ) is
large. b) at intermediate temperatures (T\TEXTsymbol{<}h)\TEXTsymbol{<}T%
\TEXTsymbol{<}T}$_{gb}${\protect\small ) all the chains in the film are in
the semi fluid state. c) at lower temperatures (T\TEXTsymbol{<}T(h)) we have
a ''sandwich structure'' with two semi fluid layers (one near each free
surface), separated by a central hard core region.\ The extrapolation
procedure of fig. 1 may lead to an apparent transition point T}$_{a}(h),$ 
{\protect\small different from T(h).}}}{\Qlb{1}}{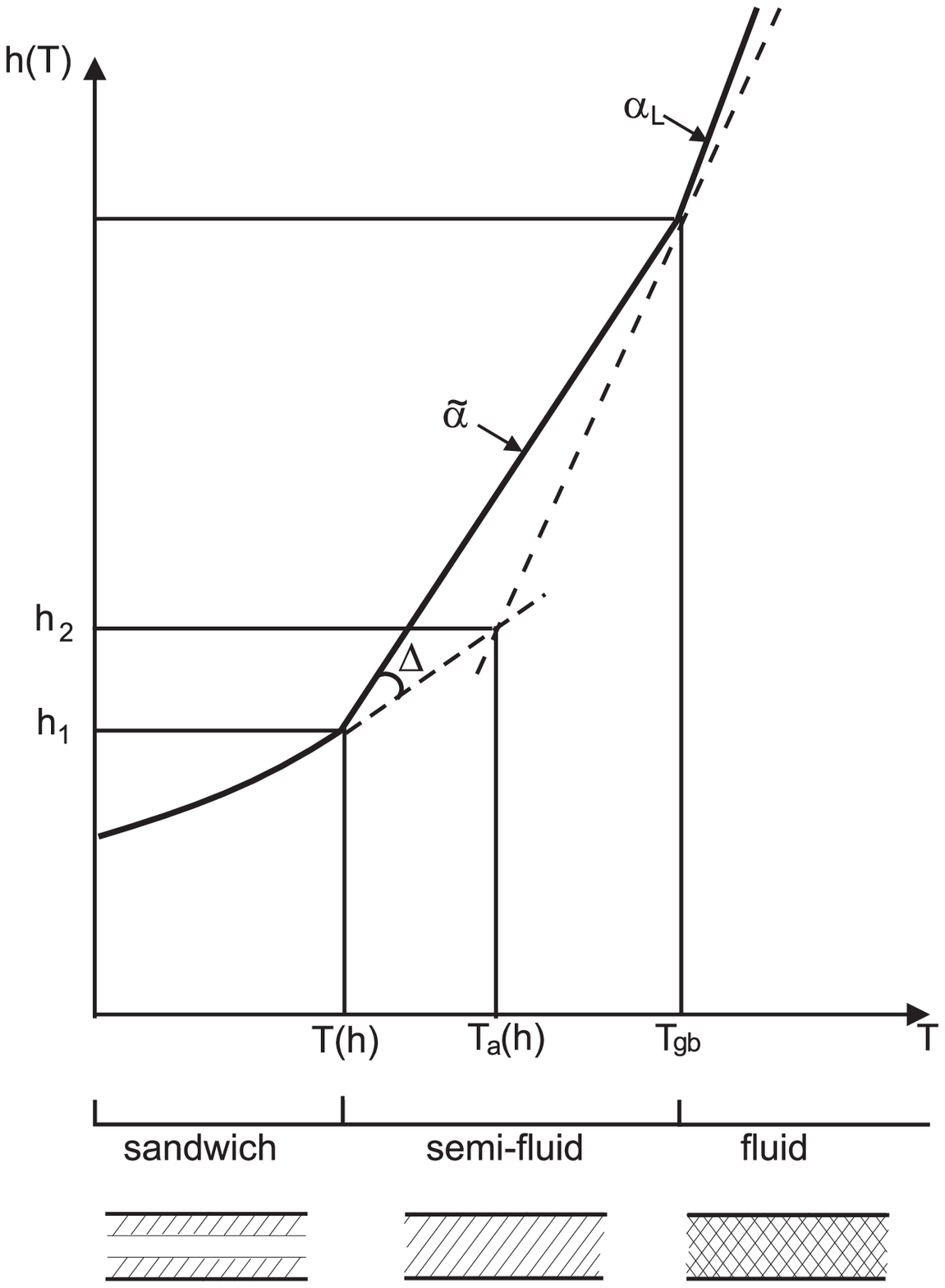}{\raisebox{-3.0536in}{\includegraphics[height=3.0536in]{fig21.eps}}}

\section{A modified version of the sliding model}

\subsection{Jump rates along an arc}

As in ref. [4], we consider an arc, or piece of chain, containing $g$
monomers between its conntact points with the surface.\ We want that a
defect (or ''kink'') be able to move all along this arc.\ The defect motions
will require certain cavity volumes $w_{1}...w_{i}...w_{g}$ for its
successive steps. The jump rates 1/$\tau _{i}$ are taken of the form:

\begin{equation}
\frac{1}{\tau _{1}}=\frac{1}{\tau _{0}}\exp \left( -\frac{w_{i}}{\text{v}(T)}%
\right)
\end{equation}

where v$(T)$ is the free volume, and is written in the simplified form:

\begin{equation}
\text{v}(T)=m(T-T_{0})
\end{equation}

($T_{0}$ is the Volger Fulcher temperature).

(Note that in eqs\ (1) and (2), the volumes $w_{i}$ and v are measured in
units of the monomer volume $a^{3}$).

We now impose that all the jump rates 1/$\tau _{i}$ along the arc, be
further than a certain rate 1/$\tau _{\exp }$ (defined by our measuring set
up).\ For the bulk systems, the cavity volume is of order $a^{3}$, and the
jump rates are of the form:

\begin{equation}
\frac{1}{\tau _{1}}=\frac{1}{\tau _{0}}\exp \left( -\frac{1}{\text{v}(T)}%
\right)
\end{equation}

The bulk glass temperature $T_{gb}$ corresponds to $\tau =\tau _{\exp }$ and
thus to:

\begin{equation}
m(T_{gb}-T_{0})=1/\ell
\end{equation}

where:

\begin{equation}
\ell \equiv \ell n\left( \frac{\tau _{\exp }}{\tau _{0}}\right)
\end{equation}

The cavity volumes $w_{i}$ required for the \textit{sliding} motion are
somewhat smaller.\ They are distributed according to a certain stochastic
law $p(w_{i)}$). If the jump at laws (i) is further than 1/$\tau _{\exp }$,
we see from eq. (1) that we must have:

\begin{equation}
\frac{w_{i}}{\text{v}(T)}<\ell
\end{equation}

or, equivalently:

\begin{equation}
w_{i}<\rho (T)
\end{equation}

where $\rho (T)$ is a reduced temperature:

\begin{equation}
\rho (T)\equiv \frac{T-T_{0}}{T_{gb}-T_{0}}
\end{equation}

The possibility of a fast jump rate ($1$/$\tau _{i}>1/\tau _{\exp }$) at
site $i$, is thus, from eq. 7:

\begin{equation}
1-Q(\rho )=\int_{0}^{p(T)}p(w_{i})dw_{i}
\end{equation}

The probability of a fast rate for \textit{all }the $g$ sites is (assuming
independence):

\begin{equation}
(1-Q)^{g}\cong e^{-Qg}
\end{equation}

(where we have anticipated the fact that $Q$ is small in the region of
interest).

From eq. (10), we see that our arc is mobile if $Qg$ is smaller than unity,
and is stuck if $Qg>1$.\ The mobility threshold corresponds to:

\begin{equation}
Q(\rho )g=1
\end{equation}

(Note that $g>>1$, and $Q$($\rho )<<1$ as announced).

\subsection{Thickness of the soft layers}

The size of the arc is related to the depth $z$ which it reaches from the
free surface. Since our chains are ideal random walks, we can put $g\sim
(z/a)^{2}$, where $a$ is again a monomer size. We thus arrive at the basic
formula for the depth of the soft layer $z$ as a function of the reduced
temperature $\rho :$

\begin{equation}
\left( \frac{z}{a}\right) ^{2}=1/Q(\rho )
\end{equation}

(This holds whenever $z$ is smaller than the coil radius $R_{0}=N^{1/2}a).$

We shall now make this more concrete by choosing the most natural form for $%
Q(\rho )$ -i.e. a gaussian tail.\ From its definition (eq. 9), we see that $%
Q(\rho =0)=1$ and $Q(\rho \rightarrow \infty )\rightarrow 0.\;$We then
postulate the following form:

\begin{equation}
Q(\rho )=\exp \left( -\frac{\rho ^{2}}{\eta ^{2}}\right)
\end{equation}

The width of the distribution of the cavity volumes $w_{i}$ is then
proportional to $\eta $. We expect $\eta $ to be somewhat smaller than
unity, because sliding motions are easier than bulk motions.\ Typically, we
shall require $\eta \sim 0.3.$\ Returning to eq.(2), we see that:

\begin{equation}
\rho =\eta \left( 2\ell n\frac{z}{a}\right) ^{1/2}
\end{equation}

The main point of interest is the cross over\ from the ''sandwich regime''
of fig.\ 2 to the single layer regime.\ This occurs when $z$ is half the
thickness of the film $z=h/2.$ The corresponding reduced temperature is:

\begin{equation}
\rho (h)=\eta \left( 2\ell n\frac{h}{2a}\right) ^{1/2}\;\;(h<2R_{0})
\end{equation}

(To use eq. 15 we must satisfy the condition $h<2R_{0}$ because the mobile
layers cannot be larger than the coil size $R_{0}$)$.$

\subsection{Saturation at large molecular weights}

The maximum allowed value for the arc length $g$ is the total length $N$ (in
monomer units).\ From eq. (10), we see that the probability of internal
mobility for the whole chain is $exp(-NQ(\rho )$).\ Let us sit at the bulk
glass temperature ($\rho =1$).\ If $N$ is too large, the sliding process
becomes non competitive ($NQ(1)>1$).\ And if it is not competitive at $%
T=T_{gb}$, it will be even less so at lower temperatures (lower $\rho $),
where $Q(\rho )>Q(1)$. Thus, we conclude that there is a maximum length:

\begin{equation}
N^{\ast }=\frac{1}{Q(1)}=\exp \frac{1}{\eta ^{2}}
\end{equation}
and a maximum depth $z^{\ast }\sim aN^{\ast 1/2}$. Thus the model does lead
to a saturation of high molecular weights.\ We do not need very small $\eta $
values to reach very high $N^{\ast }$ values.

Typically, we might choose $\eta =0.32$, leading to $N^{\ast }$ of order 1.7
x 10$^{4}$\ \ These moderate values of $\eta $ mean that the sliding process
can be important, even if it is not much easier than the bulk process.

\subsection{Discussion}

A central feature is the temperature $T(h)$ at which a film switches from
the sandwich structure to the globally semi fluid structure (fig. 2).\ A
qualitative plot of $T(h)$, deduced from eq. (15), is shown on fig. 3. For
very long chains ($N>N^{\ast }$) the whole curve is meaningful.\ For shorter
chains ($N<N^{\ast }$) the curve is valid only up to $h\cong 2R_{0}$: when $%
h>2R_{0}$, we have a sandwich structure at all temperatures below $T_{gb}.$

\FRAME{fhFU}{1.7262in}{1.1805in}{0pt}{\Qcb{{\protect\small Qualitative plot
of the temperature T(h) at which a film of thickness h switches from
''sandwich'' to ''global semi fluid''. }}}{\Qlb{3}}{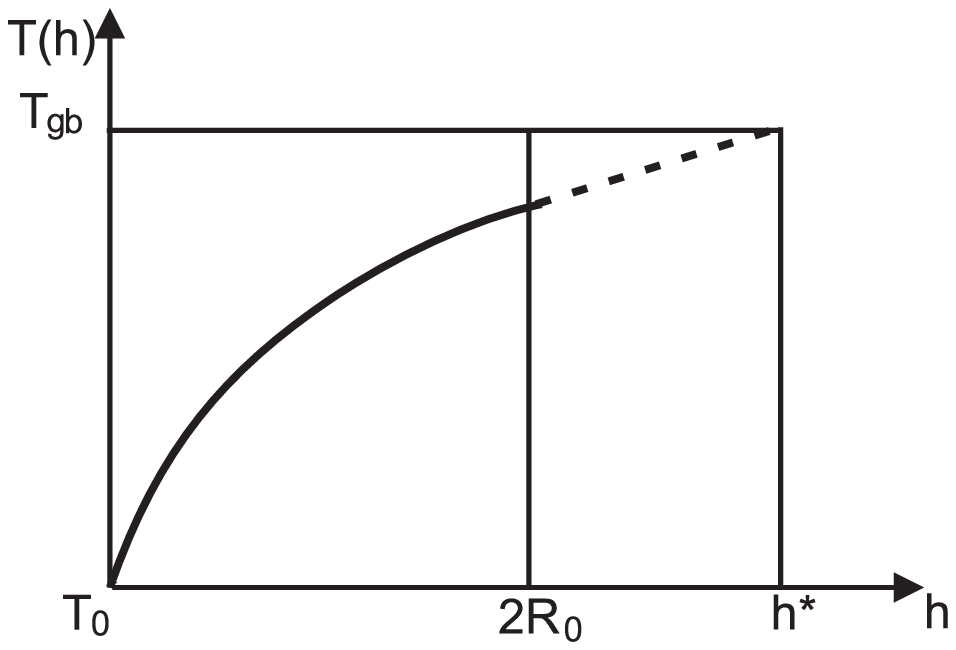}{\raisebox{-1.1805in}{\includegraphics[height=1.1805in]{fig31.eps}}}

Of course, the cut off at $h=2R_{0}$ is not sharp: even for $h=2.5R_{0}$
(for instance) there are still some chains linking the two sides.\ But,
because of the gaussian statistics of the chain conformations, the cut off
is still reasonably sharp, and meaningful.

How does the temperature $T(h)$ relate to the experiments? Initially, we
thought that the ''accident'' in the expansion plot of fig. 1 was the
signature of $T(h)$.\ On second thoughts, however, this needs not to be
true, as explained on fig. 2. The intersection scheme used to define an
accident may lead to another temperature, which we call $T_{a}(h)$.\ We
analyse $T_{a}(h)$ in the next section.

\section{The apparent glass transition \textit{T}$_{a}(h)$}

\subsection{A model of thermal expansion}

We choose a film of thickness $h<h^{\ast }$, and also impose $h<2R_{0}$, so
that the sandwich transition point $T_{a}(h)$ is meaningful.\ Our discussion
is centered on fig.\ 2.\ We are dealing with these states of polymer matter:

\qquad - the melt, with an expansion coefficient $\alpha _{L}$

\qquad - the glass, with a smaller expansion coefficient $\alpha _{g}$

\qquad - the semi fluid state with mobile arcs, for which we postulate a
third expansion coefficient $\stackrel{\sim }{\alpha }_{L}$.

This is clearly an extreme simplification: a semi fluid layer, in the
sandwich state, is really a \textit{brush} of flexible polymers rooted in
the glassy sublayer: it is strongly anisotropic and this complicate the
picture.\ But the description, in terms of one single, constant, coefficient 
$\stackrel{\sim }{\alpha }_{L}$, is a natural starting point.

In the present picture, the slopes $1/h$ $dh/dT$ (or preferably in reduced
units $1/h$ $dh/d\rho $) are constant for the melt and for a globally semi
fluid state. But the sandwich state is more delicate.\ The local slope here
is of the form:

\begin{equation}
\frac{1}{h}\frac{dh}{d\rho }=\alpha _{g}\left( 1-\frac{2z(\rho )}{h}\right) +%
\stackrel{\sim }{\alpha }_{L}\frac{2z(\rho )}{h}+\frac{\stackrel{\sim }{h}%
_{L}-h_{g}}{h}\frac{1}{h}\frac{dz}{d\rho }
\end{equation}

In eq. (7) the first two terms represent a weighted expansion for a film
with two semi fluid layers of thickness $z(\rho )$ defined by eq. (12).\ The
last term is novel.\ It involves the spontaneous thicknesses ($h_{g}$ and $%
\stackrel{\sim }{h}$) of a film which is glassy ($h_{g}$) or semi fluid ($%
\stackrel{\sim }{h}_{L}$) at the temperature of interest: when $h_{g}\neq 
\stackrel{\sim }{h}_{L}$, if the boundary between a dense phase and a less
dense phase moves, the overall thickness changes.

Of course, the parameters $h_{g}$ and $\stackrel{\sim }{h}$ are related to
the dilation coefficients $d\stackrel{\sim }{h}/d\rho =\stackrel{\sim }{%
\alpha }$, \ \ $dh_{g}/d\rho =\alpha _{g}$.\ But the difference $\stackrel{%
\sim }{h}-h_{g},$ at one chosen temperature (here at $\rho =\rho _{h}$) is
another independent parameter.

The sign of $\stackrel{\sim }{\alpha }-\alpha _{g}$ is clear: the semi fluid
phase should be more expandable upon heating than the glass phase, and $%
\stackrel{\sim }{\alpha }-\alpha _{g}>0$. But the sign of $\stackrel{\sim }{h%
}>h_{g}$ is less clear, and might depend on subtle properties of the semi
fluid brushes.

In the following discussion, we shall make two further assumptions:

\qquad (i) $\stackrel{\sim }{h}<h_{g}$.\ This is equivalent to say that, on
the plot of fig.\ 2, the slope of thickness versus temperature \textit{%
increases} when we cross $T(h)$, switching from the sandwich to the single
layer regime.

\qquad (ii) we also assume (for simplicity) that $\stackrel{\sim }{h}-h_{g}$
is temperature independent in the region of interest.

\bigskip

\subsection{Thickness/temperature plots (for $h<2R_{0}$)}

\bigskip

The temperature regions of interest are shown on fig. 2:

\qquad a) in the single layer regime, the thickness varies linearly between
a lower limit $e_{1}$ (corresponding to the sandwich transition) and an
upper limit $e_{L}$ (corresponding to the bulk glass transition point).

Thus, the difference $e_{L}-e_{1}$ is given by:

\begin{equation}
e_{L}-e_{1}=\stackrel{\sim }{\alpha }_{L}(1-\rho _{h})
\end{equation}

\qquad b) in the sandwich regime, eq. (17) holds, and the thickness
progressively increases from the glassy limit ($e=e_{S}$) up to the
transition value $e_{1}$ (at $\rho =\rho _{h}$, where $\rho _{h}$ is defined
in eq. 15). Of particular interest is the slope of $h(\rho )$, just below
the transition point, which is derived from eq. (17):

\begin{equation}
\left. \frac{dh}{d\rho }\right| _{\rho =\rho _{h}}=\stackrel{\sim }{\alpha }%
_{L}-\Delta 
\end{equation}

\bigskip

\begin{equation}
\left. \Delta =\frac{h_{g}-\stackrel{\sim }{h}}{h}\frac{1}{h}\frac{dh}{d\rho
(h)}\right| _{Z=h}=\xi /\eta ^{2}\rho h
\end{equation}

where we have used eq. (15) for $\rho (h)$ and defined a dimensionless\
parameter:

\begin{equation}
\xi =\frac{h_{g}-\stackrel{\sim }{h}}{\eta (\alpha _{L}-\stackrel{\sim }{%
\alpha }_{L})}
\end{equation}

Let us now define an \textit{apparent}$\Bbb{\ }$transition temperature $\rho
_{a}(h)$ (in reduced units) by the following construction (displayed on
fig.\ 2):

\qquad (i) we extrapolate the slope $(\stackrel{\sim }{\alpha }_{L}-\Delta )$
at $\rho >\rho (h).$

\qquad (ii) we extrapolate the fluid slope $(\stackrel{\sim }{\alpha }_{L})$
at $\rho <1.$

The intersection of these two tangents lies at $\rho =\rho _{a}$ and $%
e=e_{2.}$ Algebraically, this corresponds to:

\begin{equation}
e_{2}-e_{1}=(\alpha _{L}-\Delta )(\rho _{a}-\rho _{h})
\end{equation}

\begin{equation}
e_{L}-e_{2}=\alpha _{L}(1-\rho _{a})
\end{equation}

Eliminating $e_{2}$ between (22) and (23), we arrive at:

\begin{equation}
\rho _{a}=\frac{\rho (h)\Delta +\alpha _{L}-\stackrel{\sim }{\alpha }_{L}}{%
\Delta +\alpha _{L}-\stackrel{\sim }{\alpha }_{L}}=\frac{\rho ^{2}(h)\xi
/\eta ^{2}+1}{\rho (h)\xi /\eta ^{2}+1}
\end{equation}

The resulting relation between $\rho (h)$ and $\rho _{a}$ is shown on fig. 4:

\FRAME{fhFU}{2.4933in}{1.721in}{0pt}{\Qcb{{\protect\small Relation between
the apparent transition point }$\protect\rho _{a}(h)$ {\protect\small and
the true transition point }$\protect\rho (h)${\protect\small \ in the model
of section 3.}}}{}{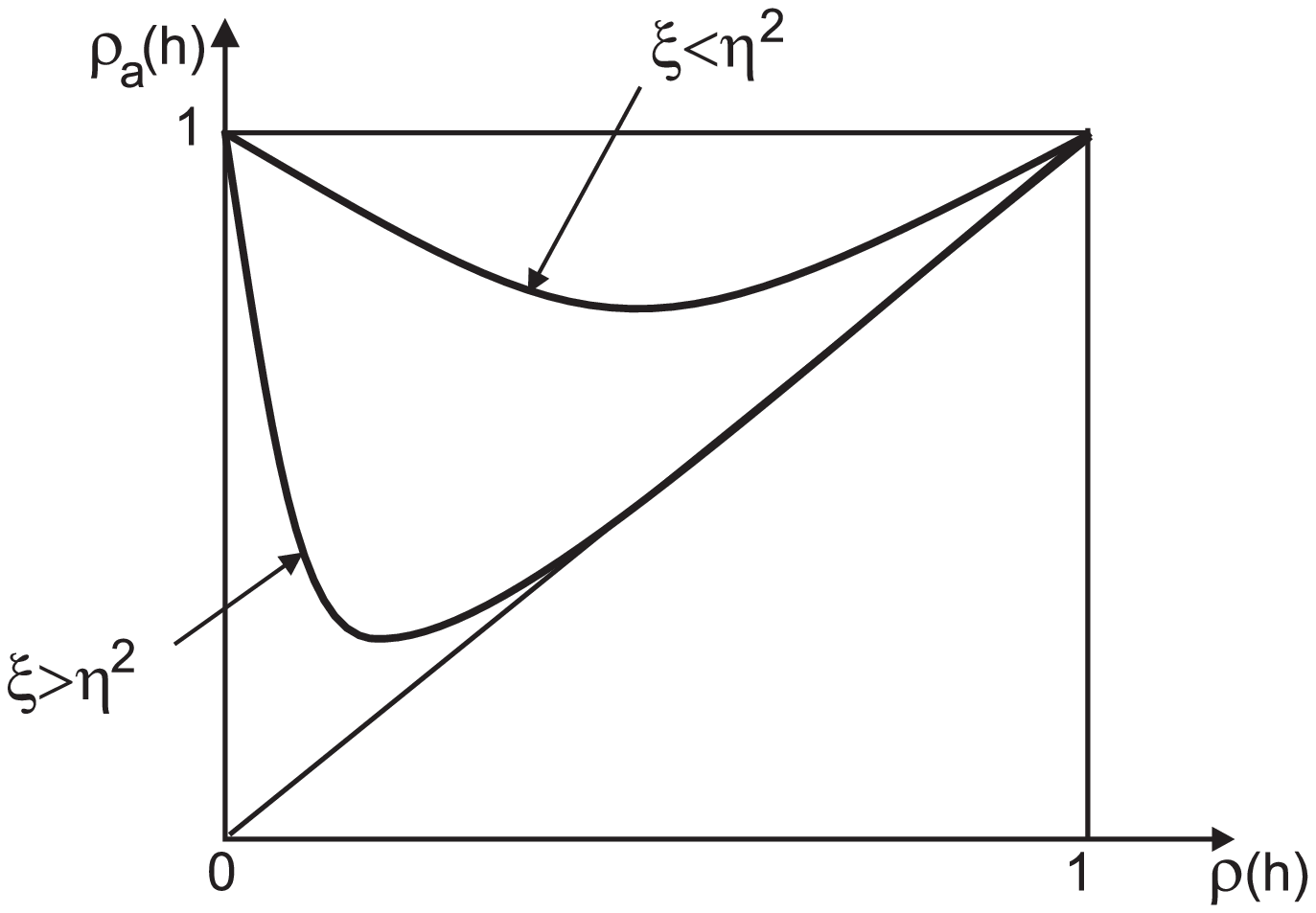}{\raisebox{-1.721in}{\includegraphics[height=1.721in]{fig4.eps}}}

\qquad a) if $\xi /\eta ^{2}<<1,$ \ \ $\rho _{a}$ is closed to unity and
much larger than $\rho (h)$ (except near $\rho =1$). Here, the distinction
between apparent and real transitions could be significant.

\qquad b) if $\xi /\eta ^{2}\stackrel{\sim }{>}1,$ \ \ $\rho _{a}$ is close
to $\rho (h)$ (except near $\rho =0$).

\bigskip

\section{Concluding remarks}

\bigskip

1) The present version of the sliding model does contain some improvements,
when compared to our precious attempts \cite{de gennes}:

\qquad a) the description of the sliding is more clear.

\qquad b) the crucial parameter is $\eta \equiv $ average cavity volume
required.for sliding/average cavity volume for bulk motion. With our
modified model, very moderate values of $\eta $ ($\sim 1/3)$ give (by eq.\
16) a limiting degree of polymerisation $N^{\ast }$ which is very high ($%
\sim 10^{4}$) and comparable to what is observed.

2) We now also realise that the apparent glass temperatures $T_{a}(h)$
derived from the thickness/temperature plots may differ from the actual
transition point $T(h)$ (where the system switches from ''sandwich'' to
''globally semi fluid). Section 3 describes one particular model for this
effect, and is only indicative. But the effect may be quite general.

3) The modified version does \textit{not }solve one major problem: why do
the drops in $T_{g}$ fade out for molecular weights smaller than $\sim 500K$ 
$(N=5000)$? Incorporating the difference between $T_{a}(h)$ and $T(h)$ does
not seem to help us here.

\bigskip

\textit{Acknowledgments}: I have greatly benefited from exchanges with J.\
Forrest and K.\ Dalnoki-Veress.\ Parts of this paper were written during a
stay at the department of chemical engineering, Universit\'{e} Laval,
Qu\'{e}bec.\ I very much wish to thank Professors R.\ Prudhomme and M.\
Bousmina for their hospitality on this occasion.

\end{document}